\documentclass{article}
\input{tcilatex}

\begin{document}

\begin{center}
The asymptotic conformal invariance in Chern-Simons theory with matter in
curved space-time

Alexandr V. Timoshkin

Tomsk Pedagogical University,634041 Tomsk,Russia
\end{center}

We consider the asymptoic behaviour of the Chern - Simons (CS) theory with
matter in curved space - time. The asymptotics of effective couplings are
discussed.

1. The study of renormalizable field theories in curved space [1,2,3,4]
proved the existance of the phenomenon of asymptotic conformal invariance.
In this short review we discuss the asymptotic behaviour\ of $3d$ CS
theories in curved space. Let us consider renormalizable abelian CS theory
with scalar and spinor in three dimensions [5]. The Lagrangian looks like:

\begin{equation}
L=\frac{1}{2}\epsilon ^{\mu \nu \lambda }A_{\mu }\partial _{\nu }A_{\lambda
}+\left\vert D_{\mu }\Phi \right\vert ^{2}+i\overline{\Psi }\widehat{D}\Psi
+\alpha \overline{\Psi }\Psi \Phi ^{\ast }\Phi -h\left( \Phi ^{\ast }\Phi
\right) ^{3}  \tag{1}
\end{equation}

Here $\ D_{\mu }=\partial _{\mu }-ieA_{\mu },\Phi ,\Psi $-complex scalar and
dirac spinor consequently, coupling constants $e$,$h$,$\alpha $ are
dimensiionless. The theory with Lagrangian (1) is multiplicatively
renormalizable.

The two - loop RG equation for coupling constants has the form [6,7]:

\[
\left( 8\pi \right) ^{2}\frac{de\left( t\right) }{dt}=0 
\]%
\begin{equation}
\left( 8\pi \right) ^{2}\frac{d\alpha \left( t\right) }{dt}=\frac{14}{3}%
\alpha ^{3}\left( t\right) -34\alpha \left( t\right) e^{4}\left( t\right)
-24e^{6}\left( t\right)  \tag{2}
\end{equation}

\[
\left( 8\pi \right) ^{2}\frac{dh\left( t\right) }{dt}=168h^{2}\left(
t\right) -84h\left( t\right) \alpha ^{2}\left( t\right) +36e^{8}\left(
t\right) +8\alpha \left( t\right) e^{6}\left( t\right) +4\alpha ^{2}\left(
t\right) e^{6}\left( t\right) -4\alpha ^{4}\left( t\right) . 
\]

It has been shown in paper [6] that for the theory with Lagrangian (1)
exists finite four cases in which the theory is finite at two - loop level:

\[
1.\ \alpha =3e^{2},\ h=-\frac{19}{4}e^{4} 
\]%
\begin{equation}
2.\ \alpha =3e^{2},\ h=e^{4};  \tag{3}
\end{equation}%
\qquad \qquad \qquad \qquad 
\[
3.\ \alpha \approx -2,23e^{2},\ h\approx -0,59e^{4}; 
\]%
\[
4.\ \alpha \approx -2,23e^{2},\ h\approx 0,62e^{4}. 
\]

In regime of finiteness the effective coupling constants are:

\begin{equation}
e\left( t\right) =e,\alpha \left( t\right) =\alpha ,h\left( t\right) =h, 
\tag{4}
\end{equation}

where values $\alpha ,h$ give one from four variants (3). Let us consider
the limit $t\longrightarrow \infty $ (infrared) for solutions of equations
(2). In this limit independently of initial values

\begin{equation}
h(t)\longrightarrow e^{4},\alpha \left( t\right) \longrightarrow 3e^{2}. 
\tag{5}
\end{equation}

Therefore the theory becomes finite and supersymmetric in the asymptotic
[8]. Consequently supersymmetry is infrared stable as far as in four
dimensions [8].Let us study now the theory in the limit $t\longrightarrow
+\infty .$ In this case we fix the initial value for $\alpha :\alpha \left(
0\right) =3e^{2}.$Then $\alpha \left( t\right) =3e^{2}$ and at $%
t\longrightarrow +\infty ,$ $h\left( t\right) \longrightarrow -\frac{19}{4}%
e^{4}$ independent of initial value $h(0)$.\ Consequently in ultraviolet
asymptotic under fixed value $\alpha \left( 0\right) =3e^{2}$ the theory
effectively becomes finite (asymptotic finitness [8]). In this case initial
value $h\left( t\right) $ is arbitrary. Fix now $\alpha \left( 0\right)
\approx -2,23e^{2}.$ Then at $t\longrightarrow \infty ,$ $h\left( t\right)
\longrightarrow 0,62e^{4}$ independently of initial value $h(0)$. The theory
again appears to be asymptotic finite. In ultraviolet limit two regims of
asymptotic finitness exist.

2. Let us consider the behavior of effective charge $\xi \left( t\right) $
in abelian CS theory with matter in curved space - time, the beta-function
on two - loop level has the form [10].

\begin{equation}
\beta _{\xi }^{\left( 2\right) }=\left( \xi -\frac{1}{8}\right) \gamma
_{m^{2}}^{\left( 2\right) },  \tag{6}
\end{equation}

where $\gamma _{m^{2}}^{\left( 2\right) }$ -gamma-function for mass of the
scalar field in two - loop approximation. Then with supposition (6), two -
loop RG equation for $\xi \left( t\right) $ has the form:

\begin{equation}
\frac{d\xi \left( t\right) }{dt}=\left( \xi \left( t\right) -\frac{1}{8}%
\right) \cdot \gamma _{m^{2}}^{\left( 2\right) }\left( t\right) ,  \tag{7}
\end{equation}

where $\gamma _{m^{2}}^{\left( 2\right) }$ is given in paper [7]. Then the
solution of equation (7) in regimes of finiteness (3) has the form:

\begin{equation}
\xi \left( t\right) =\frac{1}{8}+\left( \xi -\frac{1}{8}\right) \cdot
e^{\gamma _{m^{2}}^{\left( 2\right) }\cdot t}  \tag{8}
\end{equation}

where $\gamma _{m^{2}}^{\left( 2\right) }=0$ (for $N=2$ supersymmetric
theory) or $\gamma _{m^{2}}^{\left( 2\right) }\approx -\frac{e^{4}}{24\pi
^{2}}\cdot 11,23\times 6,77$.

Therefore at $\alpha \approx -2,23e^{2}$ the theory appears to be asymptotic
conformally invariant in $d=3$ at $t\longrightarrow \infty ,$ $\xi \left(
t\right) \longrightarrow \frac{1}{8}$ independent of initial value. At $\
t\longrightarrow -\infty ,$ $\left\vert \xi \left( t\right) \right\vert
\longrightarrow \infty $. For $N=2$ there are 18 regimes of finiteness for $%
\gamma _{m^{2}}^{\left( 2\right) }$ can be positive, negative or zero. Then
at $t\longrightarrow \infty $ the following situations are possible:

1. $\xi \left( t\right) \longrightarrow \frac{1}{8}$ (asymptotic
supersymmetry);

2. $\xi \left( t\right) =\xi $ (the asymptotic supersymmetry);

3. $\left\vert \xi \left( t\right) \right\vert \longrightarrow \infty $.

For $t\longrightarrow \infty $ the behavior of $\xi \left( t\right) $ in 1
and 3 is exchanged. Thus, it is shown that in $d3$ gauge theories also can
be asymptotic conformal invariante of exponential type, like in [9]. This
short review is mainly based on ref.[10].

Acknowledgements.

A.V.T thanks S.D. Odintsov for help at preparation this article.

References.

[1]. Buchbinder I.L., Odintsov S.D., Shapiro I.L.//Rivista Nuovo
Cim.-1989.-v.12.-p.1;
Odintsov S.D.//Fortschr. Phys.- 1991.- v.39.-p.621.

[2]. Buchbinder I.L., Odintsov S.D.//Sov. Phys. J.-1985.-2116.-p.86;
Yad.Phys.-1984.-v.40.
-p.1338; Lett. Nuovo Cim.-1985.-v.42.-p.379.

[3]. Lichtzier I.M., Odintsov S.D.//Yad.Phys.-1986.-v.47.-p.1782; Europhys.
Lett.-1988.-
v.7.-p.95; Buchbinder I.L., Odintsov S.D., Fonarev O.A.//Sov. Phys.J.-1990.-%
2116.-p.74;
Mod. Phys. Lett.-1989.-v.A4.-p.2713.

[4]. Buchbinder I.L.//Theor.Math.Phys.-1984.-v.61.-p.393; Fortschr.
Phys.-1986.-v.1986.-
v.34.-p.605.

[5]. Deser S., Jackiw R., Templeton S.//Ann. Phys.-1982.-v.140.-p.372.

[6]. Grigoryev G.V., Kazakov D.I.//Phys. Lett.-1991.-v.253B.-p.372.

[7]. Avdeev L.V., Grigoryev G.V., Kazakov D.I.//Preprint
CERN.-T.H.6091/91,1991.

[8]. Odintsov S.D., Shapiro I.L.//Lett. JETF.-1989.-v.49.-p.125;
\ Mod.PhysLett.-1989.-v.A4.-p.1479; Odintsov S.D., Shapiro I.L., Toms
D.J.//Int. J. Mod
Phys.-1991.-v.A6.-p.1829.

[9]. Buchbinder I.L., Odintsov S.D., Lichtzier I.L.//Theor. Math.
Phys.-1989.-v.79.-p.314;
Class. Grav.-1989.-v.6.-p.605; Zaripov R.Sh., Odintsov S.D.//Theor. Math.
Phys.-1990.-
v.83.-p.399; Mod. Phys. Lett.-1989.-v.A4.-p.1955.

[10] Odintsov S.D., Timoshkin A.V.//Sov. Phys. J.-1992.-2116.-P.45.

\end{document}